# A very faint core-collapse supernova in M85

**Arising from:** S. R. Kulkarni *et al. Nature* **447,** 458–460 (2007)

A. Pastorello[1], M. Della Valle[2,3,4], S. J. Smartt[1], L. Zampieri[5], S. Benetti[4,5], E. Cappellaro[4,5], P. Mazzali[6,7], F. Patat[8], S. Spiro[1,5], M. Turatto[4,5], S. Valenti[8,9]

e-mail:a.pastorello@qub.ac.uk

[1] *Astrophysics Research Centre, School of Mathematics and Physics, Queen's University Belfast, Belfast BT7 1NN, United Kingdom*

[2] *INAF Osservatorio Astronomico di Arcetri, Largo E. Fermi 5, I-50125 Firenze, Italy*

[3] *International Center for Relativistic Astrophysics Network, Piazzale della Repubblica 10, I-65122 Pescara, Italy*

[4] *Kavli Institute for Theoretical Physics, UC Santa Barbara, California 93106, USA*

[5] *INAF Osservatorio Astronomico di Padova, Vicolo dell' Osservatorio 5, I-35122 Padova, Italy*

[6] *Max-Planck-Institut für Astrophysik, Karl-Schwarzschild-Str. 1, D-85741 Garching bei München, Germany*

[7] *INAF Osservatorio Astronomico di Trieste, Via Tiepolo 11, I-34131 Trieste, Italy*

[8] *European Southern Observatory (ESO), Karl-Schwarzschild-Str. 2, D-85748 Garching bei München, Germany*

[9] *Dipartimento di Fisica, Università di Ferrara, via del Paradiso 12, I-44100 Ferrara, Italy*

**An anomalous transient in the early Hubble-type (S0) galaxy Messier 85 (M85) in the Virgo cluster was discovered by Kulkarni *et al.*[1] on 7 January 2006 that had very low luminosity (peak absolute *R*-band magnitude $M_R$ of about −12) that was constant over more than 80 days, red colour and narrow spectral lines, which seem inconsistent with those observed in any known class of transient events. Kulkarni *et al.*[1] suggest an exotic stellar merger as the possible origin. An alternative explanation is that the transient in M85 was a type II-plateau supernova of extremely low luminosity, exploding in a lenticular galaxy with residual star-forming activity. This intriguing transient might be the faintest supernova that has ever been discovered.**

Kulkarni *et al.*[1] suggest that this event (labelled M85 OT2006-1) was an anomalous luminous red nova possibly generated in a stellar merger. An overall similarity with other transients (M31 RV, V4332 Sgr and V838And) was also proposed[2], in spite of the fact that M85 OT2006-1 is significantly brighter (by two to three magnitudes). We propose an alternative scenario in which M85 OT2006-1 is an extreme case of a low-luminosity type II-plateau supernova (LL-SNIIP)[3,4], about 1.5 magnitudes fainter than the faintest LL-SNIIP discovered so far (SN1999br)[4,5].

The morphology of its light curve[1] is very similar to that of a LL-SNIIP, in terms of the plateau duration (typically 80-110 days) and the low peak luminosity ($-15 < M_V < -13$, Refs. 4-6 and Fig. 1). The spectra of M85 OT2006-1 show a red continuum and very narrow lines (H I, Ca II, Ba II, Fe II, and possibly other species, Fig. 2) which are commonly observed in LL-SNeIIP during the hydrogen envelope recombination phase[3-6]. Some features show P-Cygni profiles, others a double-component in emission. In particular, Hα shows a very narrow line (v ≈ 350 km s$^{-1}$, Ref. 1) on the top of a broader (v ≈ 800 km s$^{-1}$) component (Fig. 2, insert b) which is consistent with the typical expansion velocities of known LL-SNeIIP[4]. The properties of this subclass of SNe are explained by invoking low explosion energies (<< $10^{51}$ ergs) and very small amounts of ejected radioactive $^{56}$Ni ($\ll 10^{-2}$ M$_\odot$, where M$_\odot$ is the mass of the Sun; Ref. 4).

Core-collapse supernovae (CC-SNe) are normally observed in spirals, while M85 is a lenticular galaxy (S0). However, traces of recent episodes of star formation are observed in many local early type-galaxies, in the form of faint nebular emission lines, tidal tails, dust lines, H I gas, blue colour and radio emission. Hence the fact that M85 OT2006-1 occurred in a S0 galaxy does not categorically exclude a relatively massive progenitor star. It has been suggested that moderate-mass stars (8-12M$_\odot$) can explode as LL-SNeIIP [7-10], and it is certainly possible that E/S0 galaxies host stellar populations of such masses. Indeed, about a dozen CC-SNe have been discovered in S0 galaxies in the past five years[11]. This might well be the case for M85, which has a non-negligible radio flux (L ≈ 10$^{27}$ erg s$^{-1}$ Hz$^{-1}$), placing this galaxy in the faint tail of the radio galaxies[12]. Moreover, the strong nuclear Hα and the presence of an inner blue ring in M85 are indicative that recent (minor) episodes of star formation occurred[13].

One of the arguments advanced by Kulkarni *et al.*[1] against a massive star origin for M85 OT2006-1 was that no star was visible in deep pre-explosion Hubble Space Telescope images. However the detection limit of $m_{F850LP}$ = 24.7 (as quoted by Kulkarni *et al.*[1]) is not sufficient to exclude a moderately massive precursor. If it was a type K to M supergiant, after adopting the assumptions on the distance and the reddening by Kulkarni *et al.*[1], a luminosity limit of logL/L$_\odot$ ≈ 4.8 is obtained. Using current stellar evolutionary tracks for an isolated red supergiant, we obtain an upper mass limit of 15M$_\odot$. The deep detection limit in the blue filter ($m_{F475W}$ = 26.8) can be used to restrict blue supergiants (B-type, as in the case of SN1987A) to less than about 12M$_\odot$. Small star clusters of initial mass about 300M$_\odot$ (obtained using the stellar population models of S*tarburst99*; Ref. 14) would be undetected with these magnitude limits. Assuming a typical initial mass function in such a cluster, one could quite reasonably expect a few stars above the limit of $8M_\odot$. In light of all of this, we suggest a plausible physical interpretation in which the M85 transient is an extremely faint CC-SN, characterized by a very low explosion energy ($5\text{-}10\times10^{49}$ ergs), a small amount of $^{56}$Ni ($<< 10^{-3} M_\odot$) and a total ejected mass of 6-9$M_\odot$.

**Figure 1 | Absolute R-band magnitude light curves for M85 OT2006-1 and LL-SNeIIP.** The data for the SN sample, adopted explosion epochs, interstellar extinction and distance moduli (calibrated with $H_0 = 72$ km s$^{-1}$ Mpc$^{-1}$) are from Refs. 4 and 6. Unpublished data of the faint SNIIP 2003Z, with distance modulus (m - M) = 32.64 and interstellar extinction E(B - V) = 0.04, are also displayed. For M85 OT2006-1, we make use of a direct estimate of the surface brightness fluctuation distance to M85 (Ref. 15), giving (m - M) = 31.33 ± 0.14 (18.5 ± 1.2 Mpc), adopting the same interstellar absorption estimate ($A_R = 0.4$) as in Kulkarni *et al.*[1]. With an average plateau magnitude R = 19.65, we obtain an absolute magnitude of $M_R = -12.1$. LL-SNeIIP cover an extended range of luminosity, and thus the light curve of M85 OT2006-1 (considering the photometric errors from Kulkarni *et al.*[1]) is consistent with that of an extremely faint SNIIP.

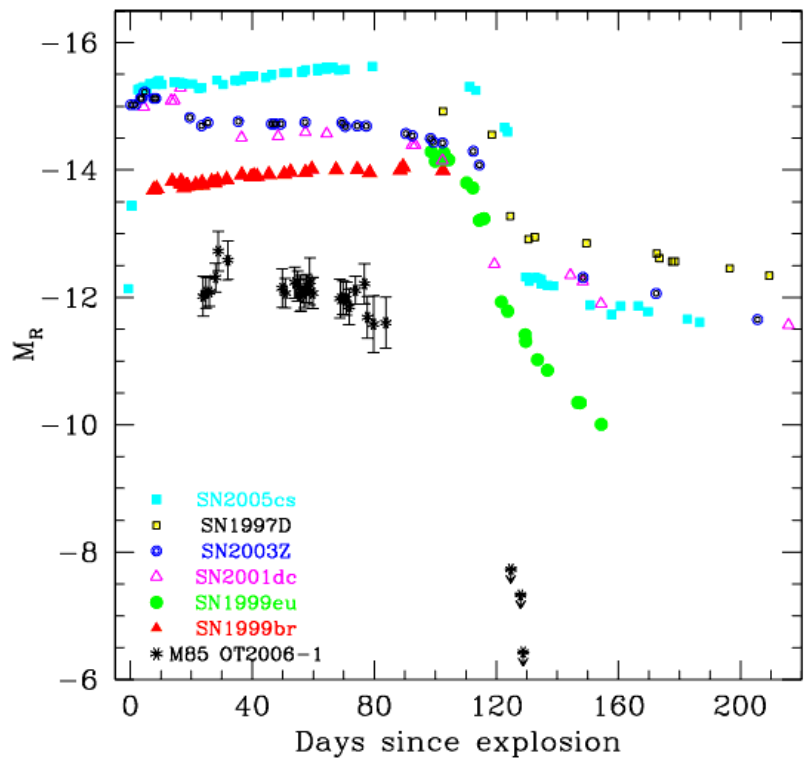

**Figure 2 | Spectral properties of M85 OT2006-1 and LL-SNeIIP.** The spectrum of M85 OT2006-1 obtained on 24 February 2006 (from Ref. 1, kindly provided by S. R. Kulkarni and A. Rau, de-reddened by $A_R = 0.4$ magnitudes) is compared with those of SN1999br and SN1999eu observed at the end of the plateau[4]. The position of the H Balmer lines, Ba II $\lambda 6497$ and the near infrared Ca II triplet are marked by vertical dashed lines. These lines are visible in the spectrum of the transient, while there is no clear evidence of molecular TiO bands, characterising spectra of red giants and the peculiar luminous red novae proposed in refs 1 and 2. $F$ is the flux (in erg s$^{-1}$) scaled to an arbitrary constant. The axis units are the same for **a** and **b**. **a,** the region of the P-Cygni Ba II $\lambda 6497$ + H$\alpha$ lines in SN1999eu and M85 OT2006-1. **b,** Three-component fit (broader P-Cygni feature with $V_{exp} \approx 700\text{-}800$ km s$^{-1}$, plus narrow emission) to the H$\alpha$ profile. The red line shows the level of the continuum, the green lines show separately the three components used to fit the line, and the dotted line is the resulting three-component fit.

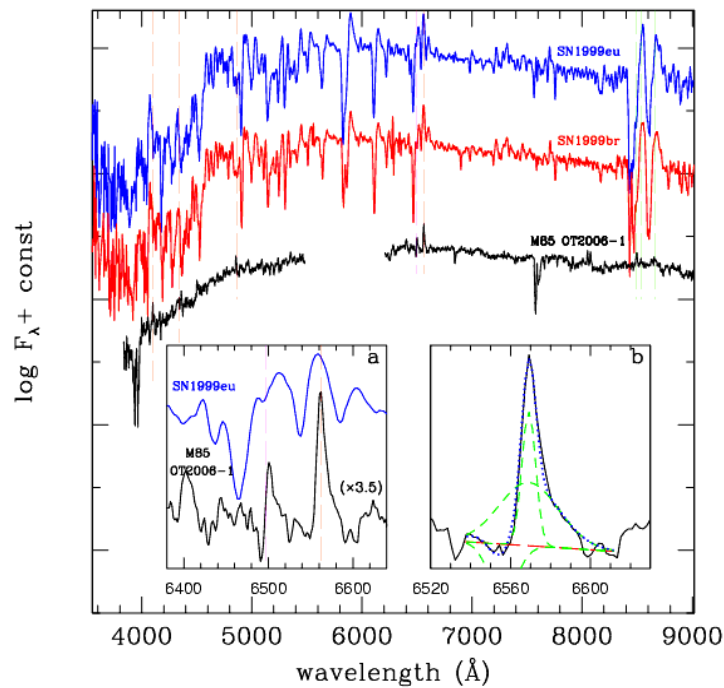